\documentstyle[aps,preprint]{revtex}
\begin{document}

\begin{center}
{\Large Nuclear shape transition at finite temperature in a
relativistic mean field approach}\\

\vskip 1 cm
{B. K. Agrawal$^{1}$, Tapas Sil$^{2}$, J. N. De$^{2}$, 
     and S. K. Samaddar$^{1}$}\\
\vskip 1 cm

$^{(1)}$ Saha Institute of Nuclear Physics, 1/AF,\\
Bidhannagar, Calcutta - 700064, India\\
$^{(2)}$ Variable Energy Cyclotron Centre, 1/AF,\\
Bidhannagar, Calcutta - 700064, India
\end{center}

\begin{abstract}
The relativistic Hartree-BCS theory is applied to study the 
temperature dependence of nuclear shape and pairing gap
for $^{166}Er$ and $^{170}Er$.  
For both the nuclei, we find that as temperature increases the
pairing gap vanishes leading to phase transition
from superfluid to normal phase as is observed in nonrelativistic
calculation. The deformation evolves from prolate shapes to
spherical shapes at $T\sim 2.7$ MeV. 
Comparison of our results  for heat capacity with the ones
obtained in the  non-relativistic mean field framework indicates that in
the relativistic mean field theory the shape transition 
occurs at a temperature about 0.9 MeV higher and is relatively weaker.
The effect of thermal shape fluctuations on the temperature
dependence of deformation is also  studied.
Relevant results for the level density parameter are further presented.

\vskip .5 cm

PACS numbers: 21.10.Ma, 21.60.-n, 27.70.+q
\end{abstract}

\vskip .5 cm

{\it Keywords:} Relativistic mean field, Hot nuclei, Shape transition,  
Level density parameter

\newpage
\section{Introduction}
The relativistic mean field (RMF) theory \cite{ser,gam,rin} has been very
successful in describing  the ground state (zero temperature)
properties of nuclei over the entire periodic table
in recent years. The binding energies,
charge radii and the ground state deformations are reproduced very well;
the charge distributions also compare extremely well with the experimental
data. This theory has  proved to be very fruitful in explaining 
\cite{vre,pos,lal} various details of exotic nuclei near the drip lines.
In contrast to the nonrelativistic models, the RMF theory uses a single set
of parameters to explain all these properties.
To our knowledge, this approach has not yet been exploited to
understand the properties of hot nuclei except some preliminary investigations
for closed-shell nuclei \cite{gam1}. 
The response of nuclear shapes to thermal
excitations, for example, has been experimentally studied from the
shapes of the giant dipole resonances (GDR) built on excited states
\cite{sno,gaa}. Theoretically such finite temperature effects have been
studied till date in the nonrelativistic framework such as the 
Hartree-Fock-Bogoliubov (HFB) theory \cite{goo1,goo2} and the Landau
theory of phase transition \cite{alha1,alha2}. Such theories
qualitatively explain, for example, the temperature evolution of
nuclear shapes. Recent experiments indicate, however, that a quantitative
estimate of the persistence of the ground state deformation \cite{kel}
with temperature may be missing in some cases.

 Against this backdrop, we undertake the study of the thermal response
to nuclear properties, particularly the deformation and the level
density in the RMF framework. The HFB calculations employ a model
hamiltonian in a limited model space with pairing-plus-quadrupole
interaction. This may not be very realistic once temperature comes 
into play. Moreover, to make the calculations numerically tractable,
an inert core is assumed. This may be questionable at moderately
high temperatures. The RMF approach we use is effectively free from these
limitations. The model space employed is sufficiently large 
and all the nucleons are treated on equal footing.

 In the present work, we employ the nonlinear $\sigma -
\omega -\rho$ version of the RMF theory \cite{gam}. 
In absence of a simple relativistic prescription
for the pairing interaction, it is introduced phenomenologically.
We take two
representative systems, namely, $^{166}Er$ and $^{170}Er$. 
We investigate the thermal evolution of the nuclear shapes
and the pairing gaps. 
The temperature
dependence of the specific heat as a possible signature of phase
transition in pairing and nuclear shapes is explored.
The temperature variation  of the nuclear level density
parameter which has a pre-eminent role in understanding nuclear reactions
is further studied. Effects of thermal fluctuations of the nuclear
shapes on the deformation and the nature of the phase transition
is also discussed.

   The theoretical framework used is briefly discussed in Sec.II.
The results and discussions are presented in Sec.III and the Sec.IV
contains the concluding remarks.

\section{Formalism}
The  Lagrangian density for the nucleon-meson  many-body system \cite{gam}
is taken as
\begin{eqnarray}
\label{lag}
{\cal L}&=& \bar\Psi_i\left ( i\gamma^\mu \partial_\mu - M\right )\Psi_i
+ \frac{1}{2} \partial^\mu\sigma\partial_\mu\sigma - U(\sigma)
- g_\sigma \bar\Psi_i \sigma\Psi_i\nonumber\\
&& - \frac{1}{4}\Omega^{\mu\nu}\Omega_{\mu\nu}
+\frac{1}{2}m_\omega^2\omega^\mu \omega_\mu - g_\omega \bar\Psi_i \gamma^\mu
\omega_\mu\Psi_i 
-\frac{1}{4}\vec{R}^{\mu\nu} \vec{R}_{\mu\nu} + \frac{1}{2} m_\rho^2 \vec{\rho}^\mu\vec{\rho}_\mu\nonumber\\
&&- g_\rho \bar\Psi_i \gamma^\mu\vec{\rho}_\mu\vec{\tau}\Psi_i
-\frac{1}{4}F^{\mu\nu}F_{\mu\nu} - e\bar\Psi_i \gamma^\mu \frac{(1-\tau_3)}{2} A_\mu\Psi_i
.
\label{Lag}
\end{eqnarray}
The meson fields included are the isoscalar $\sigma$ meson, 
the isoscalar-vector $\omega$
meson and the isovector-vector $\rho$ meson. The arrows in eq. (\ref{Lag})
denote the isovector quantities.  The Largrangian contains  a nonlinear 
scalar self-interaction term $U(\sigma)$ of the $\sigma$ meson: 
\begin{equation}
U(\sigma) = \frac{1}{2}m_\sigma^2 \sigma^2 + \frac{1}{3}g_2 \sigma^3 + 
\frac{1}{4}g_3\sigma^4
.
\end{equation}
This term is important  for appropriate description of the
surface properties \cite{bog}. The quantities 
$M$, $m_\sigma$, $m_\omega$ and $m_\rho$ are the nucleon, $\sigma$, 
$\omega$ and the $\rho-$meson masses, respectively, while 
$g_\sigma$, $g_\omega$, $g_\rho$ and $e^2/4\pi = 1/137$ are
the corresponding coupling constants for the mesons and the photon. 
The field tensors of the
vector mesons and of the electromagnetic fields have the following structure:
\begin{eqnarray}
\Omega^{\mu\nu} & = & \partial^\mu\omega ^\nu - \partial^\nu\omega^\mu,\\
{\vec {\bf R}}^{\mu\nu} & = & \partial^\mu {\bf{\vec \rho}}^\nu 
- \partial^\nu{\bf {\vec \rho}}^\mu - g_\rho(\vec\rho^\mu\times\vec\rho^\nu),\\
F^{\mu\nu}& = & \partial^\mu A^\nu - \partial^\nu A^\mu.
\end{eqnarray}

The variational  principle gives  the equations of motion. 
The mean field approximation is introduced at this stage by treating the fields as $c-$numbers or classical
fields. This results  in a set of coupled equations, namely the Dirac equation with potential
terms for the nucleons and the Klein-Gordon type equations with sources  for the
mesons and the photon. For the static case, along with
the time reversal invariance and charge conservation the equations get simplified.
The resulting equations, known as RMF equations, have the following form.
The Dirac equation  for the nucleon is  
\begin{equation}
\label{dir}
\{-i{\bf\alpha}\cdot{\bf \nabla} + V({\bf r}) +\beta\left [M+S({\bf r})
\right ] \} \Psi_i = \epsilon_i\Psi_i,
\end{equation}
where $V({\bf r})$ represents  the {\it vector } potential 
\begin{equation}
V({\bf r}) = g_\omega \omega_0({\bf r})  + g_\rho \tau_3{\bf \rho}_0({\bf r})
+ e\frac{(1 - \tau_3)}{2} A_0({\bf r}),
\end{equation}
and $S({\bf r})$ is the {\it scalar} potential 
\begin{equation}
S({\bf r}) = g_\sigma\sigma({\bf r}),
\end{equation}
which contributes  to the effective mass as 
\begin{equation}
M^*({\bf r}) = M + S({\bf r})
.
\end{equation}

The Klein-Gordon equations for the mesons and the electromagnetic fields  with the
nucleon densities as sources  are 
\begin{eqnarray}
\label{sig}
\left \{ -\Delta + m_\sigma^2\right \} \sigma({\bf r}) & = & -g_\sigma\rho_s({\bf r})
-g_2\sigma^2({\bf r}) - g_3\sigma^3({\bf r}),\\
\label{ome}
\left \{ -\Delta + m_\omega^2\right \} \omega_0({\bf r}) 
& = & g_\omega \rho_v({\bf r}),\\
\label{rho}
\left \{-\Delta +  m_\rho^2\right \} \rho_0({\bf r})  & = & g_\rho\rho_3({\bf r}),\\
\label{pho}
-\Delta A_0({\bf r}) & = & e\rho_c({\bf r}).
\end{eqnarray}
The corresponding densities are  
\begin{eqnarray}
\rho_s & = & \sum_i n_i\bar\Psi_i \Psi_i,\nonumber\\
\rho_v & = & \sum_i n_i\Psi^\dagger_i \Psi_i,\nonumber\\
\rho_3 & = & \sum_i n_i\Psi^\dagger_i\tau_3 \Psi_i,\nonumber\\
\rho_c & = & \sum_i n_i\Psi^\dagger_i\frac{(1-\tau_3)}{2} \Psi_i.
\end{eqnarray}
Here the sums are taken over the particle states only, i.e., 
the negative-energy states are neglected. 
The partial occupancies ($n_i$)
at finite temperature in the constant pairing gap approximation (BCS) is 
\begin{equation}
n_i = \frac{1}{2}\left [1-\frac{\epsilon_i-\lambda}{\tilde\epsilon_i}
\left ( 1 - 2f(\tilde\epsilon_i,T)\right )\right ],
\end{equation}
with 
$f(\tilde\epsilon_i,T) = 1/(1 + e^{\tilde\epsilon_i/T})$;
$\tilde\epsilon_i = \sqrt{(\epsilon_i - \lambda)^2 + \Delta^2}$
is the qusiparticle energy 
 where $\epsilon_i$ 
is the single-particle energy for the state $i$.  The chemical potential
$\lambda$
for protons (neutrons) is obtained from the requirement
\begin{equation}
\sum_i n_i = Z \> (\> N)
\end{equation}
The sum is taken over proton (neutron) states.
The gap parameter $\Delta$ is obtained by minimising the free energy
\begin{equation}
F = E - TS,
\end{equation}
where
\begin{equation}
\label{ene}
E(T) = \sum_i \epsilon_i n_i + E_\sigma + E_{\sigma NL}+E_\omega + E_\rho 
+ E_C +E_{pair} + E_{CM} - AM,
\end{equation}
and 
\begin{equation}
S = -\sum_i \left [ f_i lnf_i + (1-f_i) ln(1-f_i)\right ],
\end{equation}
with
\begin{eqnarray}
\label{esi}
E_\sigma& = &-\frac{1}{2}g_\sigma \int d^3r \rho_s({\bf r}) 
\sigma({\bf r}),\\
\label{esinl}
E_{\sigma NL}& = &-\frac{1}{2} \int d^3r
\left\{\frac{1}{3}g_2 \sigma^3({\bf r})+\frac{1}
{2}g_3\sigma^4({\bf r})\right\},\\
\label{emo}
E_\omega& = &-\frac{1}{2}g_\omega\int d^3r \rho_v({\bf r})\omega^0({\bf r}),\\
\label{erh}
E_\rho& = &-\frac{1}{2}g_\rho\int d^3 r \rho_3({\bf r}) \rho^0({\bf r}),\\
\label{ecou}
E_C& = &-\frac{e^2}{8\pi}\int d^3r \rho_C({\bf r}) A^0({\bf r}),\\
\label{epair}
E_{pair}& = &-\frac{\Delta^2}{G},\\
\label{ecm}
E_{CM}& = &-\frac{3}{4}\hbar\omega_0 = -\frac{3}{4}41 A^{-1/3}.
\end{eqnarray}
Here $G$ and $A$ are the pairing strength and the mass number respectively.
The single-particle energies  and the fields appearing in eqs. (\ref{ene}) 
- (\ref{erh}) are obtained  from the self-consistent solution of 
eqs. (\ref{dir}) - (\ref{rho}). We generate these
self-consistent solutions using the well tested basis expansion method
as described in detail in Refs. \cite{gam,rin1}.

The self-consistent solutions for the RMF equations
at a given temperature is obtained by minimizing the free energy $F$
which yields the equilibrium or the most probable value of the 
quadrupole deformation ($\beta_2^0$) and
the proton (neutron) pairing gaps $\Delta_p$ ($\Delta_n$). To incorporate the
effects of the thermal fluctuations, 
a constrained calculation is performed to generate the free energies
for the deformations away from the equilibrium value. 
Average value of a quantity then
can be obtained as 
\begin{equation}
\label{average}
\bar O = \frac{\int d\beta_2 O(\beta_2)
e^{-\Delta F(\beta_2)/T}}
{\int d\beta_2 e^{-\Delta F(\beta_2)/T}}
\end{equation}
where $\Delta F=F(\beta_2)-F(\beta_2^0)$ and
$O(\beta_2)$ is the expectation value of the operator
$\hat O$ at a fixed value of $\beta_2$ and $T$.
The quantity $e^{-\Delta F/T}$ is a measure of the probablity 
for the nucleus to 
have deformation $\beta_2$ at the temperature $T$.

\section{Results and Discussions}

We have chosen $^{166}Er$ and $^{170}Er$ as two representative
systems for our calculations. 
The results presented  are obtained using the
NLSH parameter set \cite{rin1} for  the values of the coupling 
constants and  masses for the
mesons and nucleons.  The pairing strength $G$ is taken to be
$29/A$ and $21/A$ for protons and neutrons, respectively;
they reasonably reproduce the observed pairing gaps  at zero temperature
obtained from odd-even mass differences.
The values of the chemical potential and the pairing gap  
are determined using  all the single particle states up to
$2\hbar\omega_0$ above the Fermi surface without 
assuming any core, i.e., the mean field solution  is generated  by taking into
account all the nucleons in the systems considered.

    At finite temperatures, the nucleus is not strictly in thermal
equilibrium. In order to treat the system as an isolated one in
equilibrium, one has to take into account corrections due to excitations
of nucleons in the continuum. This is generally done through a 
subtraction procedure, by treating the liquid plus vapour phase together
and then the vapour phase seperately \cite{bon}. However,
calculations of the nuclear level density parameter at finite temperature
show that the results are insensitive \cite{agr1} to the continuum 
corrections for $T$ upto $\sim$ 3 MeV; we have therefore not included the 
effects due to the continuum in the present calculations as temperatures
above it are not relevant. 

  In subsec.3.1, the results for the most probable values (i.e, the mean
field values ) of the different observables are presented. With
increasing temperature, thermal fluctions build up which may shift the
average values from the most probable ones. In subsec.3.2, we discuss 
the results with the inclusion of thermal fluctuations. Fluctuations in
both the $\beta_2$ and $\gamma$- degrees of freedom of the nuclear
shape should be incorporated, however, because of simplicity and
computational economy, we have included only the $\beta_2$- fluctuations
in the present calculations. 

\subsection{Mean field results}

  We have calculated the most probable values of the quadrupole 
deformation parameter, neutron and proton pairing gaps, heat capacity
and the level density parameters for the aforesaid systems as a function
of temperature. 

In Fig. 1, the variation of  the quadrupole
deformation parameter $\beta_2$ as a function of temperature
is shown. The ground state deformations obtained here are almost
close to the ones calculated in Refs.\cite{goo1,goo2} using
nonrelativistic mean field theory with pairing plus quadrupole
(P+Q) interaction. The nature of the temperature dependence
of the deformation in the two cases are also not very different. 
However, whereas in the nonrelativistic case the phase transition
from prolate to spherical shape occurs at $T\sim$1.8 MeV, in the
present case the said transition is found at a higher temperature,
$T\sim$2.7 MeV.
The signature of the phase-transition can be inferred
by examining other observables like heat capacity which we consider later.

In Fig. 2, the results for the evolution of the neutron and proton
pairing gaps with temperature are displayed. For both the nuclei  the
pairing gaps monotonically decrease with the increase in temperature.
It is seen that for both the nuclei, the neutron pairing gaps
vanish almost at the same temperature $T\sim$0.4 MeV; similarly
the proton pairing gaps vanish at $T\sim$0.45 MeV. These results
are not much in variance with those obtained in the nonrelativistic 
P+Q model. 

The vanishing of the nuclear deformation and the pairing gaps 
with temperature 
indicates that there is a shape
transition from prolate to spherical and also a transition from
the superfluid phase to the normal phase. 
To understand the nature 
of the transitions, the heat capacity at various temperatures are calculated.
The specific heat $C$ for a given temperature is obtained using 
\begin{equation}
C(T) = \frac{\partial E^*}{\partial T}
\end{equation}
where $E^*$ is the excitation energy of the nucleus. 
In Fig. 3, the  temperature variation of the
specific heat for $^{166}Er$ and $^{170}Er$ nuclei is plotted.
At $T \sim$ 0.4 MeV, two closely seperated peaks in the $C(T)$ curves
are seen. They correspond to the dissolution of the neutron and proton
pairing gaps.  These are 
the characteristic signatures of second order phase transitions
from superfluid to normal phase. 
It is further noted that the heat capacities for the two systems
have broad bumps at $T\sim 2.7$ MeV.
This signifies a weak second order phase transition corresponding
to the transition of the nuclear shape.
This result is at variance with those obtained in the P+Q model
or in the nonrelativistic
calculation \cite{alha1}
based on the Landau theory of phase transitions where
a strong second order phase transition is observed. 

We now  present the results  for the temperature dependent level
density parameter $a$. The parameter $a$ can be obtained using the
excitation energy and the entropy  as follows,
\begin{eqnarray}
\label{ae}
E^* & = & aT^2,\\
\label{as}
S &  = & 2aT
.
\end{eqnarray}
The parameter $a$ obtained using above eqs. (\ref{ae}) and (\ref{as})
would be equal  provided it is independent of temperature
\cite{agr}. In Fig. 4,   the inverse
level density parameter $K=A/a$ ($A$ is the mass number of the nucleus)
is plotted as a function of 
temperature. The subscripts $E$ and $S$ are used
to distinguish the two definitions given by
eqs. (\ref{ae}) and (\ref{as}), respectively.  At lower
temperature,  both $K_E$ and $K_S$  shoot up
due to the collapse of the pairing gaps. At higher temperatures, 
$K_E$ and $K_S$ are quite
close to each other and there is no appreciable variations in their values
in the temperature range considered.
This is due to the weak transition of the nuclear shapes  
from deformed to spherical at   $T \sim 2.7$ MeV.

In order to test the  sensitivity of the results presented above to
the choice of the parameter set  and the model space, we have repeated  the
calculations for the nucleus $^{166}Er$  with  the NL3  parameter 
set \cite{lal2,lal3}.
In comparison to the results for NLSH parameter set it is found that
the NL3 parameter set gives  slightly larger ($\sim 10\%$)
value of  the ground state deformation. For $T> 1.5\> MeV$, 
the values of deformation obtained in both  NLSH and
NL3 parameter sets  are almost identical. The temperature evolution of the 
pairing gaps and level density parameters  are practically the same
with the two parameter sets.
We extend the model space to include single particle 
states upto $3\hbar \omega_0$
above the Fermi surface instead of $2\hbar \omega_0$ as used above.
For this extended model space the pairing strength $G$ is
adjusted to reproduce the ground state pairing gap.  We do not
find significant changes  in the values of deformation parameter and 
the pairing energy in the temperature range of interest ($T\sim 3$ MeV).  
In the extended model space, the value of the inverse level density
parameter is also nearly the same. 
The model space used is thus found to be sufficient for our calculations.
To estimate  the importance of the continuum corrections to the inverse
level density parameter we calculated the occupancy, $n^{(+)}$, of the
single-particle states with positive energy. For $T\> < \> 1$ 
MeV practically there is no
particle in the positive energy states ($n^{(+)} = 0$) and at 
the highest temperature of interest studied  
$n^{(+)}/A = 0.018$ which is very small. Thus it is expected that
the continuum corrections may not play an important role in 
the temperature range we study. The continuum effects may grow stronger
for $T>3$ MeV, however, this is beyond the transition temperatures
and so we have not taken this into account.

The shape transition temperature and the inverse level density parameter
obtained in the present model are higher compared to those calculated
in the (P+Q) model \cite{goo1,goo2}. Moreover, in the later model, the 
deformation falls to zero sharply whereas in the present case, this
fall is comparatively a little slower. The origin of these differences
can be attributed to the single-particle level spectra. 
In Fig.5, 
the energy levels for the different proton orbitals around
the Fermi surface for the nucleus $^{170}Er$ at $T$=3.0 MeV are displayed.
Also the proton single-particle energies in the (P+Q) model 
are shown at this temperature. In both models the orbits
are spherical at this temperature. The striking difference
between the orbitals in the two models is the large energy gap across
the Fermi surface in the RMF model. Similar is the case for neutrons.
These tend to reduce the entropy and the excitation energy and hence
the level density parameter. 
The increase in the shape transition temperature (by $\sim$ 0.9 MeV)
in the RMF model as compared to the P+Q model may be traced back to the 
specific level structure and the gap across the Fermi surface in the 
two models and their thermal response.

\subsection{Results with thermal fluctuations}

Experimental data  on  giant dipole resonances built on excited states
give estimates of the average values of nuclear deformation at 
finite temperatures. They can be calculated 
by taking into account the thermal fluctuation effects around 
the 'most probable' value. Nuclei being finite
systems, one expects  that  thermal  fluctuations  would play an
important role  in the quantitative  estimation of  various nuclear 
observables at  finite temperatures. In fact, it has
been shown in Ref\cite{alha3} that  the experimental data on GDR built on 
the excited states can be explained reasonably only if the effects from the
thermal fluctuations of all the quadrupole degrees of freedom for
nuclear shape are included. In the following, we
consider the thermal fluctuations only in the $\beta_2$ degree of freedom.
Inclusion of fluctuations in the $\gamma$-degrees of freedom is
very involved and computer-intensive and therefore not considered presently.
We perform  constrained calculations for free 
energy at fixed deformations $\beta_2$. 
The free energy surface shows multiple minima for temperatures below
the shape transition temperature. The average value of $\beta_2$ is
calculated using eq.(27). In Fig.6, we have displayed the plots of 
free energy surface $\Delta F$ for $^{170}Er$ at a few temperatures
ranging from 2.0 to 3.0 MeV where $\Delta F$ is measured from the 
lowest minimum (prolate in the present case) at the corresponding
temperature.
For $T$ = 2.0 MeV,  the free energy surface has
minima at $\beta_2= -0.16$ and 0.23. The free energy difference 
between these two minima is 0.70 MeV (i.e. $\Delta F/T=0.35$).
As the temperature approaches the shape transition temperature
($T\sim$2.7 MeV), 
both the minima  in the free energy surface  merge and flatten the
bottom part of the free energy surface. This is evident from the
middle panel of Fig.6 corresponding to $T=$ 2.65 MeV. 
With  further increase  in temperature,   the free energy 
surface with single
minimum  at zero deformation broadens.
The variation of the
average deformation ($\bar \beta_2$)as a function of temperature 
obtained through eq.(\ref{average}) for $^{170}Er$ is shown in Fig.7.
The average value of $\beta_2$ differs
significantly from its most probable value (see Fig. 1) for temperatures
 above 1 MeV. The reason
for such differences can be understood as follows. The free energy surface
for $^{170}Er$ at  low temperatures ($T < 2$ MeV) shows 
two well seperated minima.
One of these minima lies at  prolate deformation ($F_p$) 
and the other one is
at oblate deformation ($F_o$).
Their difference $\Delta F^0 = F_o - F_p$
as a function of temperature is shown in Fig.8.
For $T>0.5$ MeV, $\Delta F^0$
decreases monotonically to zero.
The initial rise of $\Delta F^0$ can be 
attributed to the quenching of the pairing correlations; it is seen
that without pairing, $\Delta F^0$ decreases uniformly with temperature. 
The decrease in $\Delta F^0$ enhances the probability  of finding
the nucleus with oblate deformation and hence the average deformation
is smaller than the most probable one for temperatures below the 
transition temperature. The  deviation of the
average  deformation  from the most probable value is also governed by 
$\Delta\beta_2 = \beta_2^p-\beta_2^o$. 
Here, $\beta_2^p$ and $\beta_2^o$ are the
deformations corresponding to the prolate and the oblate minima.
With rise in temperature, they come closer and ultimately
merge as shown in Fig. 9. The persistence of $\bar \beta_2$
beyond the transition temperature is attributed to the asymmetry
in the free energy surface as seen in the bottom panel of Fig.6.
A similar behavior for $^{166}Er$ is
also seen.

 In the mean field approximation, the thermal evolution of the most
probable value of the deformation is seen to be associated with two
phase transitions, the first being a superfluid phase to normal
phase transition at $T\sim$ 0.4 MeV and the other a weak second
order phase transition in the shape at $T \sim$ 2.7 MeV. It would
be interesting to see how thermal fluctuations affect these phase
transitions. In Fig. 10, the specific heat for the system $^{170}Er$
is displayed as a function of temperature with inclusion of 
fluctuations. At low temperatures for $T < 1.0$ MeV, there is no
perceptible change in the specific heat and thus the transition
from superfluid to normal phase is not affected by fluctuations.
At a little higher temperature, however, the broad bump seen in
Fig.3 at $\sim $ 2.7 MeV is seen to be shifted to $\sim$ 1.6 MeV 
and becomes also a little sharper. The phase transition in nuclear
shape thus occurs at
a lower temperature and is not as weak as seen in the mean field
approximation. Normally, one would expect the phase transition 
to get diluted due to presence of fluctuations. Here one sees the
opposite. The evolution of the free energy and the energy profiles
with temperature showing two minima (one prolate and the other
oblate) in deformation space is responsible for this observation.
Here the role of thermal fluctuations becomes significant only
in the intermediate temperature domain (temperature in the 
range 1 - 2.5 MeV, approximately) due to decrease in $\Delta
F^0$ and $\Delta \beta_2$ (see Figs. 8 and 9). The delicate
interplay of these quantities in the averaging augurs a change
in the heat capacity signalling a somewhat sharper shape transition
at a lower temperature compared to the most probable one.  

The effect of thermal fluctuations of the shape coordinate $\beta_2$
on the inverse level density parameter $K$ for the system $^{170}Er$
is shown in Fig.11. The full line and the dashed line correspond to
results of our calculations for $K_S$ and $K_E$, respectively. The
corresponding lines with circles represent those with inclusion of
the fluctuations. It is seen that the fluctuations shift the minima
for $K_S$ and $K_E$ from 2.7 MeV to 1.6 MeV, the corresponding
transition temperature. Beyond the transition temperature, the
upward slope for $K$ is also found to be higher with fluctuations
included. Similar is the behavior for $^{166}Er$.

\section{Conclusions}

The relativistic mean-field approach together 
with pairing effects
included through the constant gap approximation  has been applied to study
some properties of axially deformed $^{166}Er$ and $^{170}Er$ 
nuclei at finite temperature. The temperature dependence of the quadrupole
shape for both the nuclei are found to be practically 
the same; similar is the case
for the pairing gaps. The deformation remains close to its 
ground state value ($\beta_2\sim$ 0.3) for $T$ less than 1.5 MeV and
falls to zero quite sharply at $T\sim$2.7 MeV.
Pairing gaps vanish at $T$ =0.4 - 0.5 MeV leading to
transition from a superfluid phase to normal phase. Compared
with the results obtained in the nonrelativistic mean field
framework, the relativistic calculations yield vanishing pairing 
gaps at practically the same temperature and the fall is a bit slower.
The evolution of the heat capacity with temperature and the transition
from superfluid to normal phase do not behave very differently. The 
nature of the shape transition obtained
in the two approaches are also not too different except that in
the present case, the approach of the deformation to zero
is relatively slower. 
This  slower fall results in a  broad bump
in the specific heat around the shape transition temperature 
($T \sim $ 2.7 MeV).

   Fluctuations are expected to influence the phase transitions.
To explore these aspects, we have taken into consideration thermal
fluctuations in the $\beta_2-$ degrees of freedom. The effects of 
fluctuations are found to be imperceptible below $T \sim$ 1 MeV
and therefore the pairing transitions are not affected. However, the 
influence on the shape transitions is found to be very significant;
the transition temperature drops down from $\sim $ 2.7 MeV to 
$\sim$ 1.6 MeV with inclusion of thermal fluctuations. It also
becomes a little sharper. The specific nature of the free energy
surface in the deformation space and its evolution with temperature
are responsible for such a behavior. Fluctuations in the
$\gamma-$ degrees of freedom would also have a role to play in the
phase transitions mentioned. We have not included these presently
as they are computationally too intensive.

The authors gratefully acknowledge Prof. P. Ring for providing
them with the computer code to generate RMF
solutions for axially deformed nuclei at zero-temperature.

\newpage
\noindent{\bf Figure Captions:}

\begin{itemize}
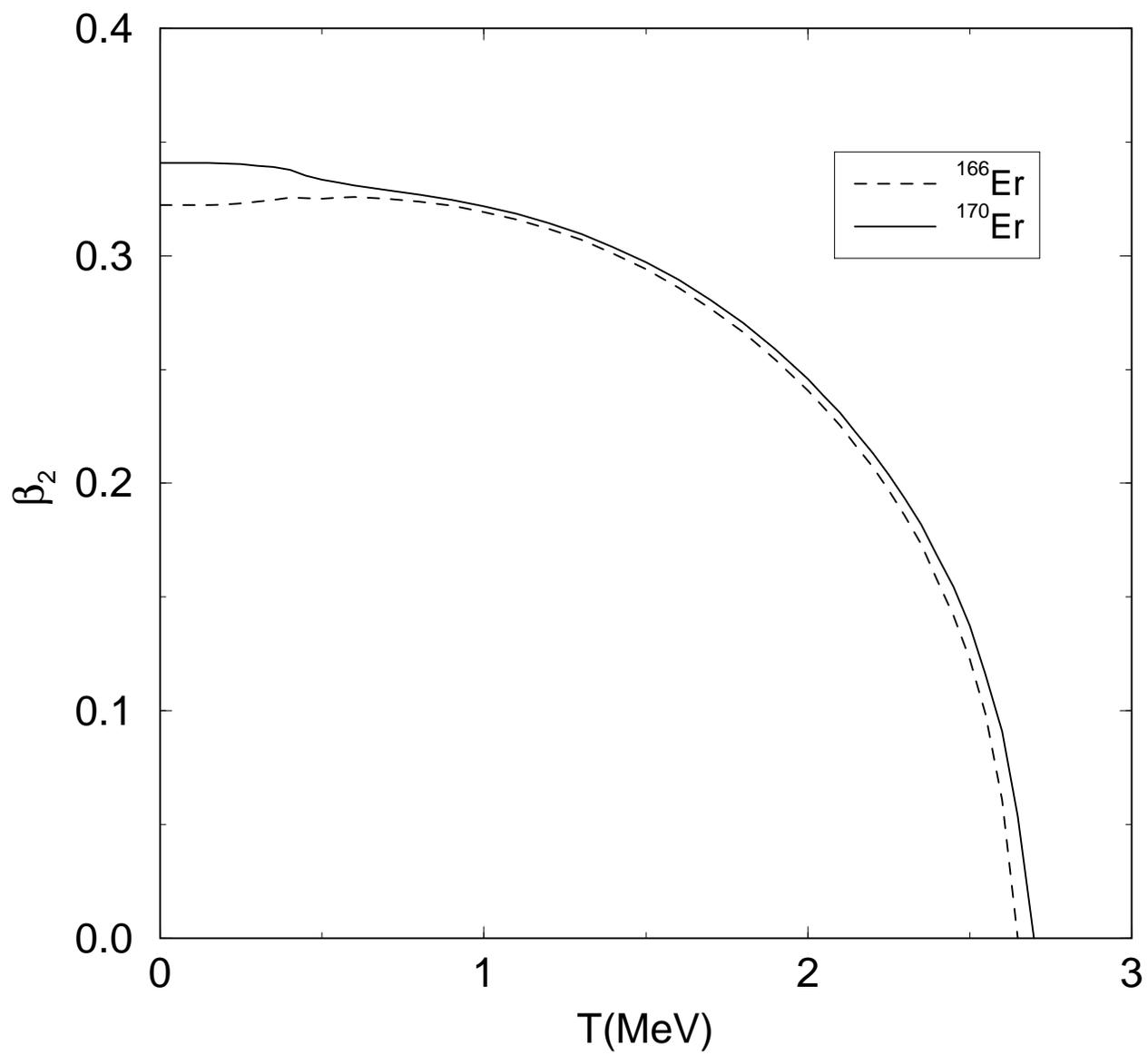
\item[Fig. 1:] Variation of the most probable quadrupole 
deformation $\beta_2$ as a function of
temperature for $^{166}Er$ and $^{170}Er$. 
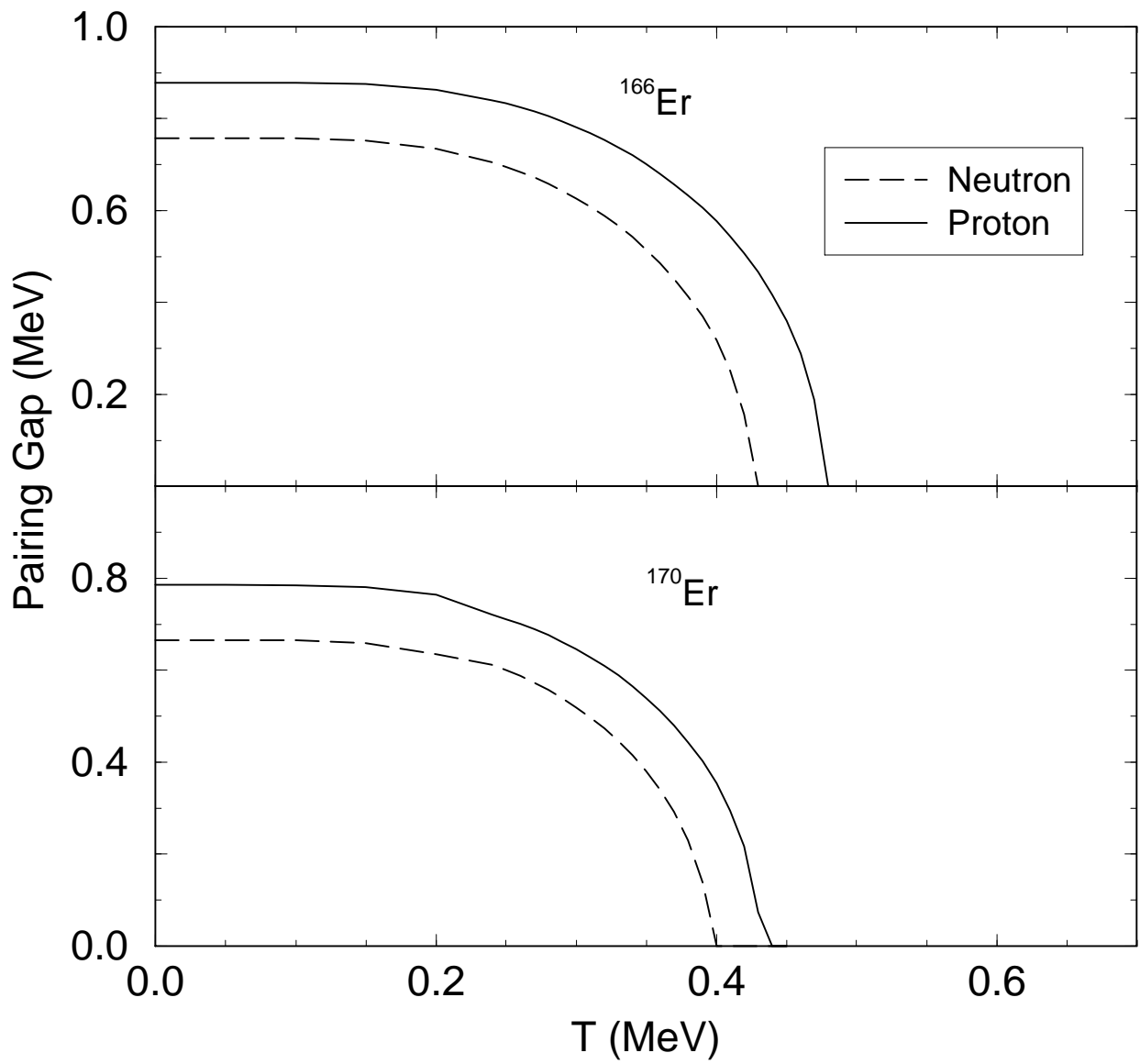
\item[Fig. 2:] Temperature evolution of neutron and proton pairing gaps for
$^{166}Er$ and $^{170}Er$.
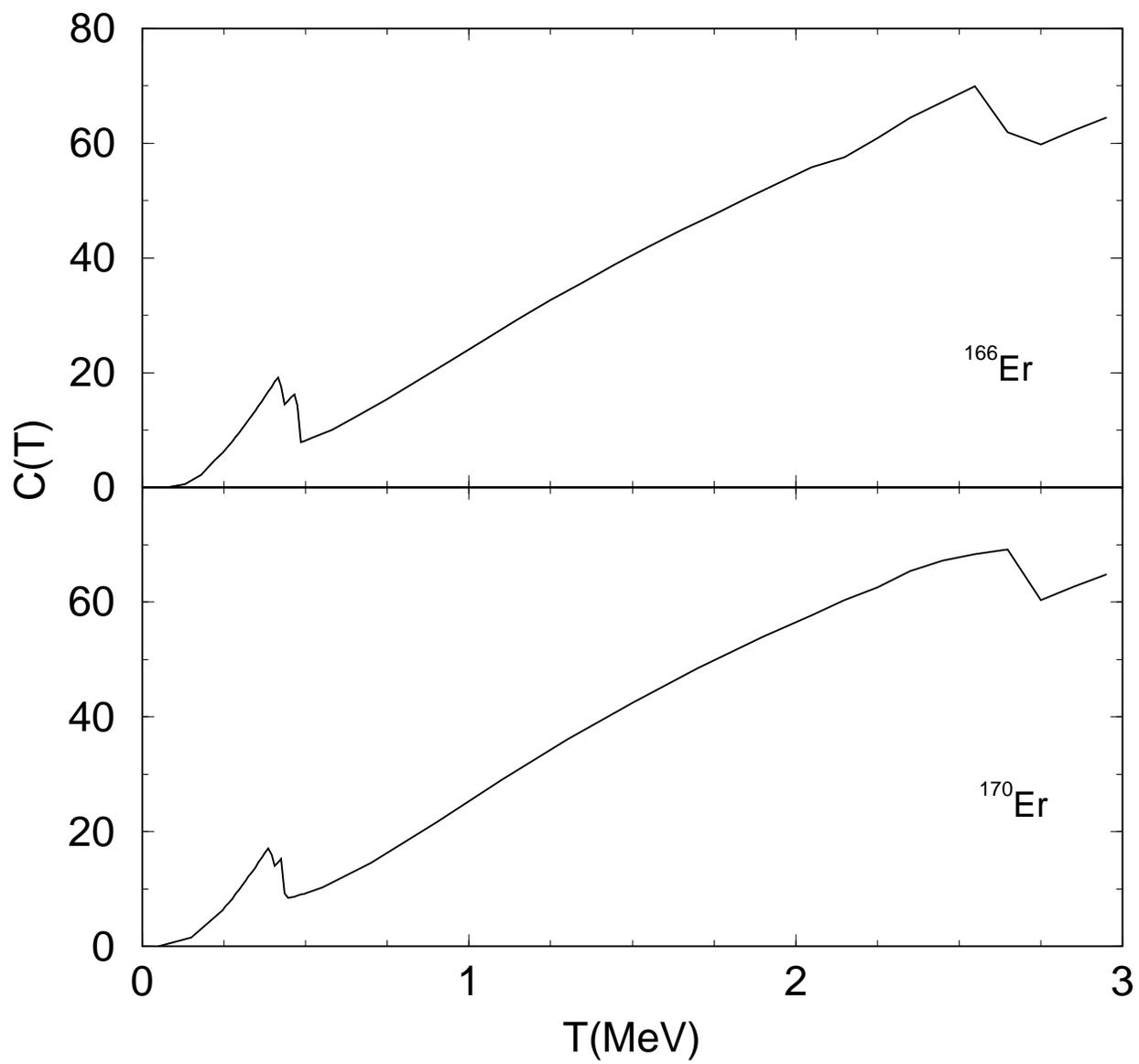
\item[Fig. 3:] Variation of specific heat as a function
of temperature for $^{166}Er$ and $^{170}Er$.
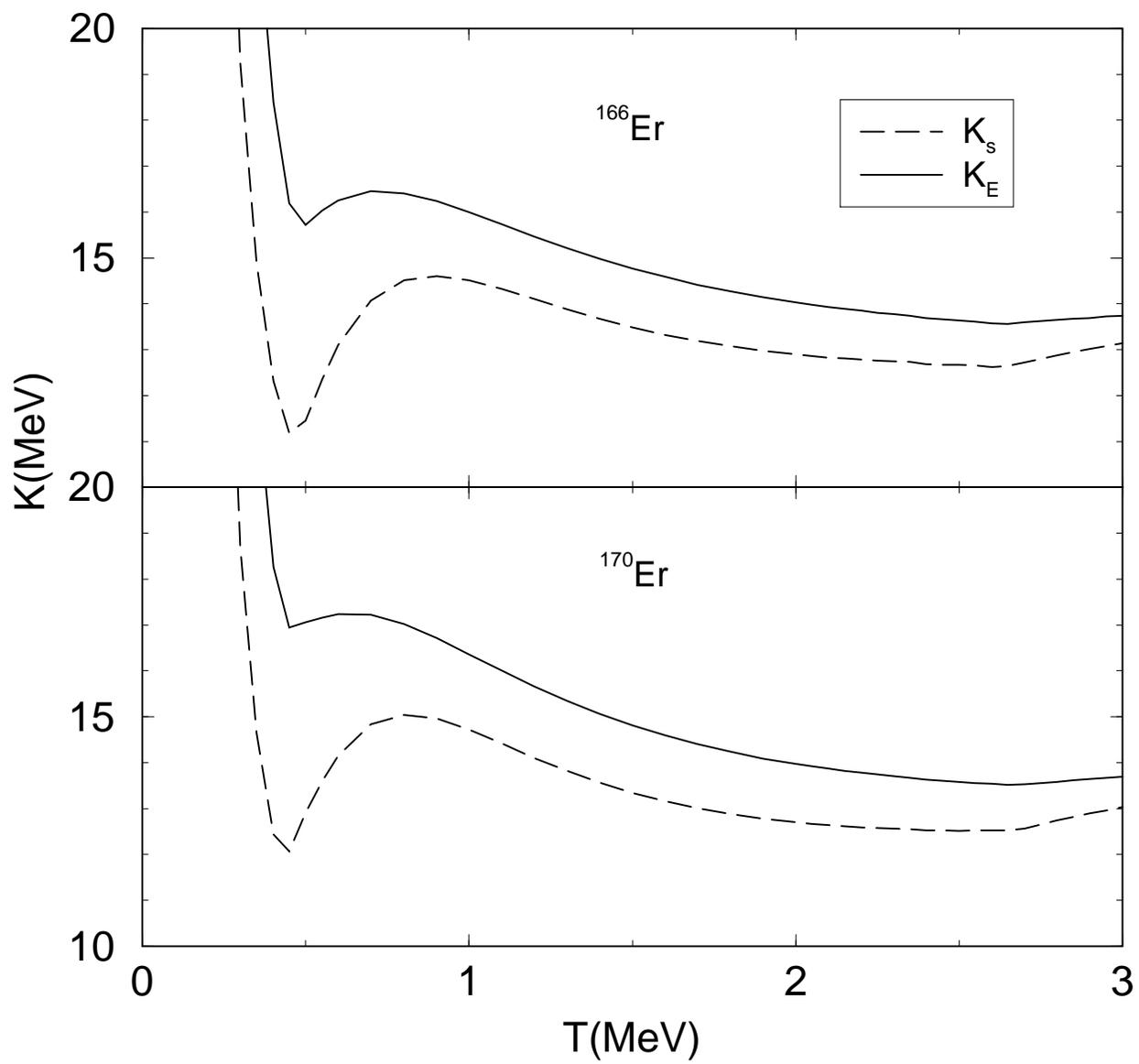
\item[Fig. 4:] Temperature  dependence of inverse level density parameters
$K_E$ and $K_S$ determined from the excitation energy and the entropy,
respectively.
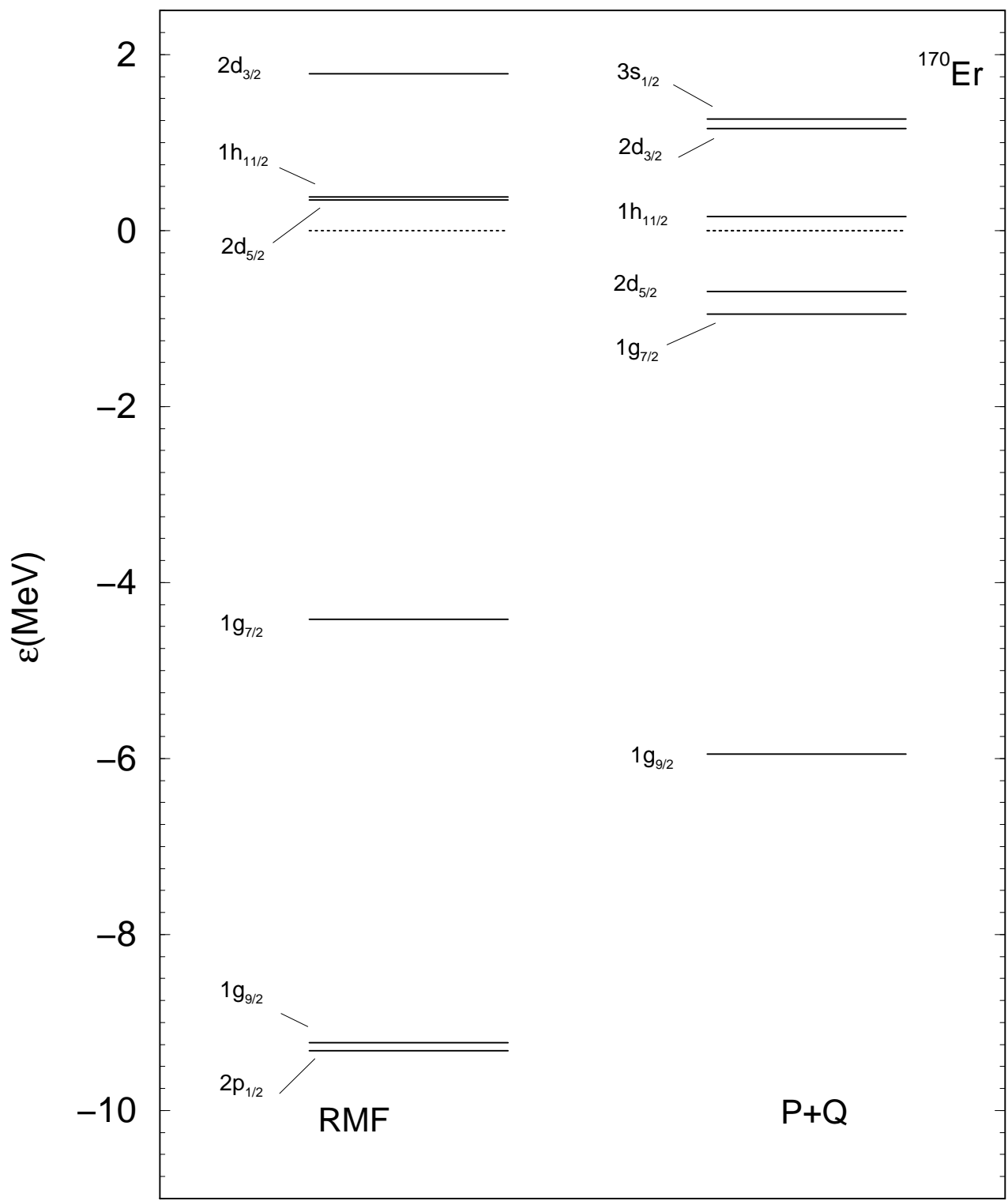
\item[Fig. 5:] Spectra of the proton single-particle energies 
around the Fermi surface for
$^{170}Er$ at $T = 3.0$ MeV in the RMF and the (P+Q) models. 
The single-particle energies $\epsilon$ are
shown relative to the respective chemical potentials.
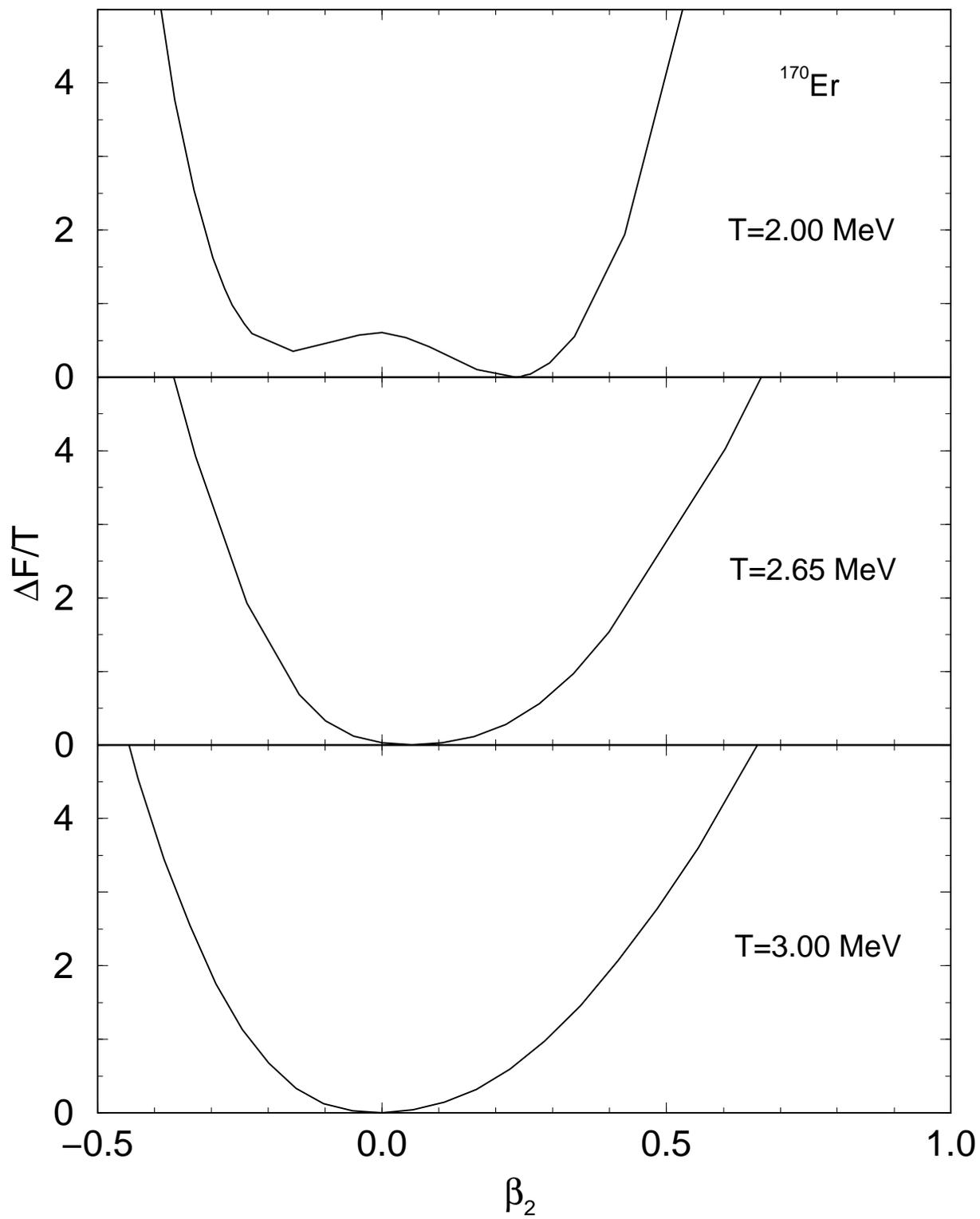
\item[Fig. 6:] Plots for the free energy surface at various 
temperatures for $^{170}Er$. The free energies at different temperatures 
are measured from the respective lowest minimum.
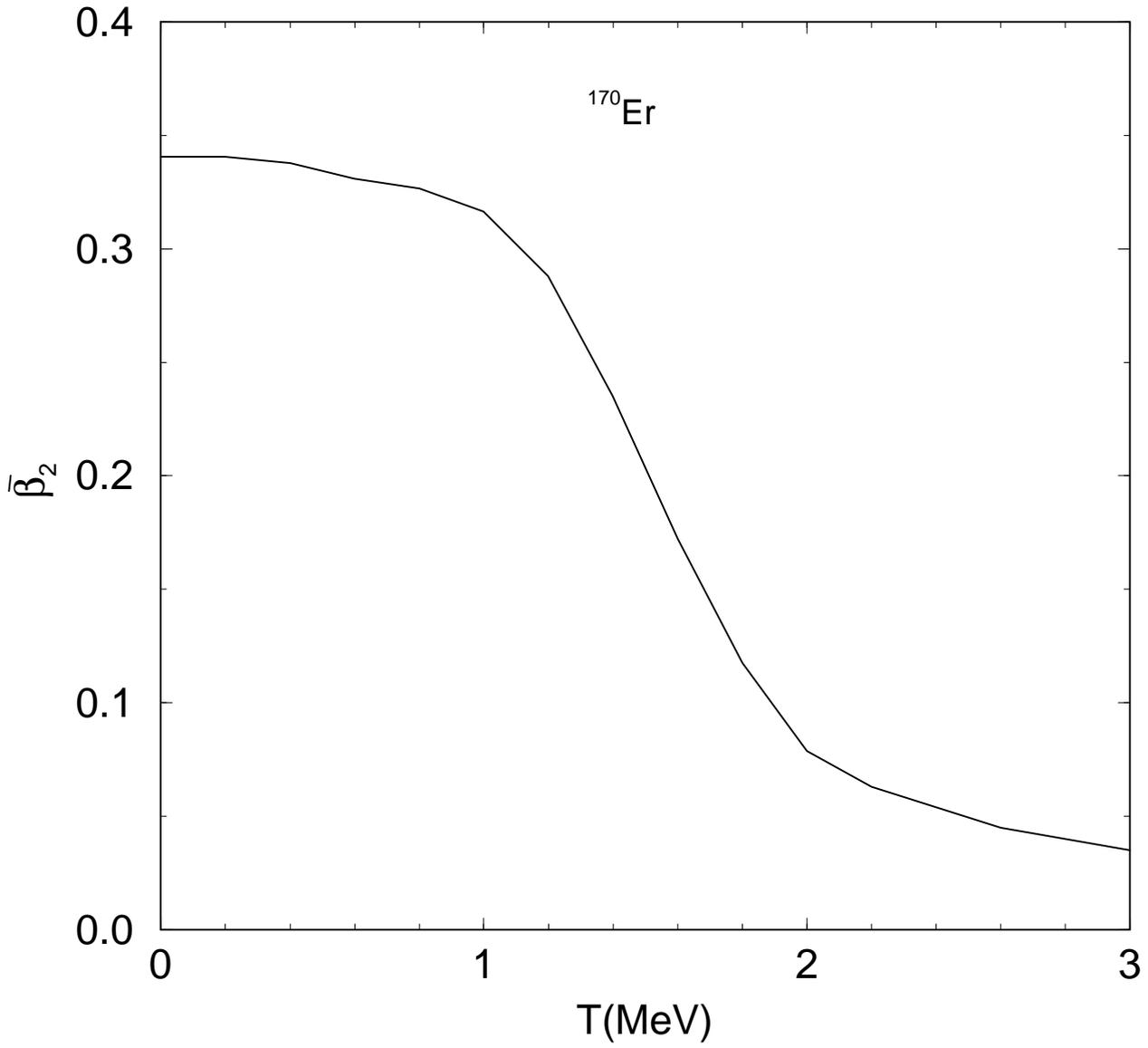
\item[Fig. 7:] Average value of quadrupole deformation  as a function
of temperature for $^{170}Er$.
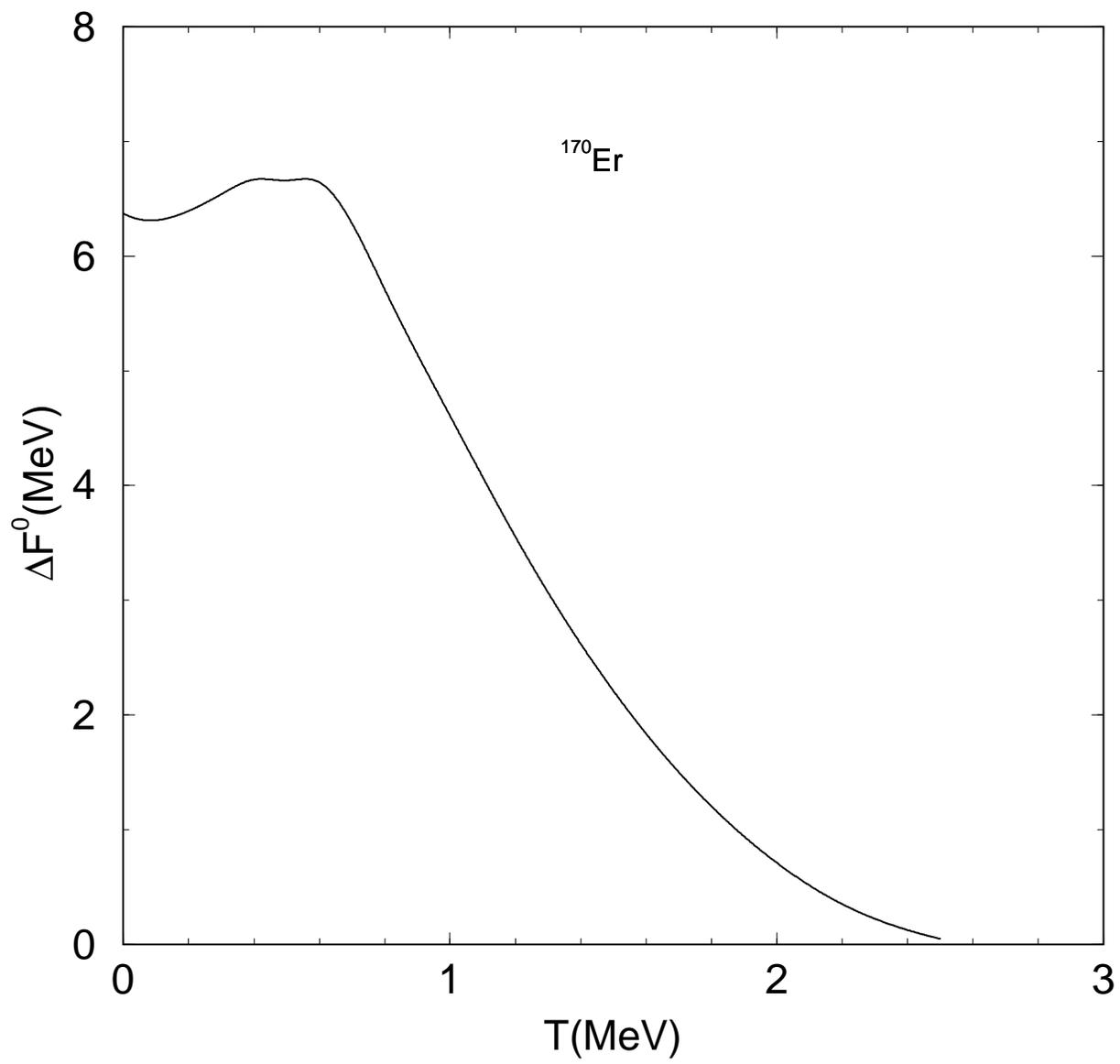
\item[Fig. 8:] The free energy  difference between the prolate and the 
oblate minima as a function of temperature for $^{170}Er$.
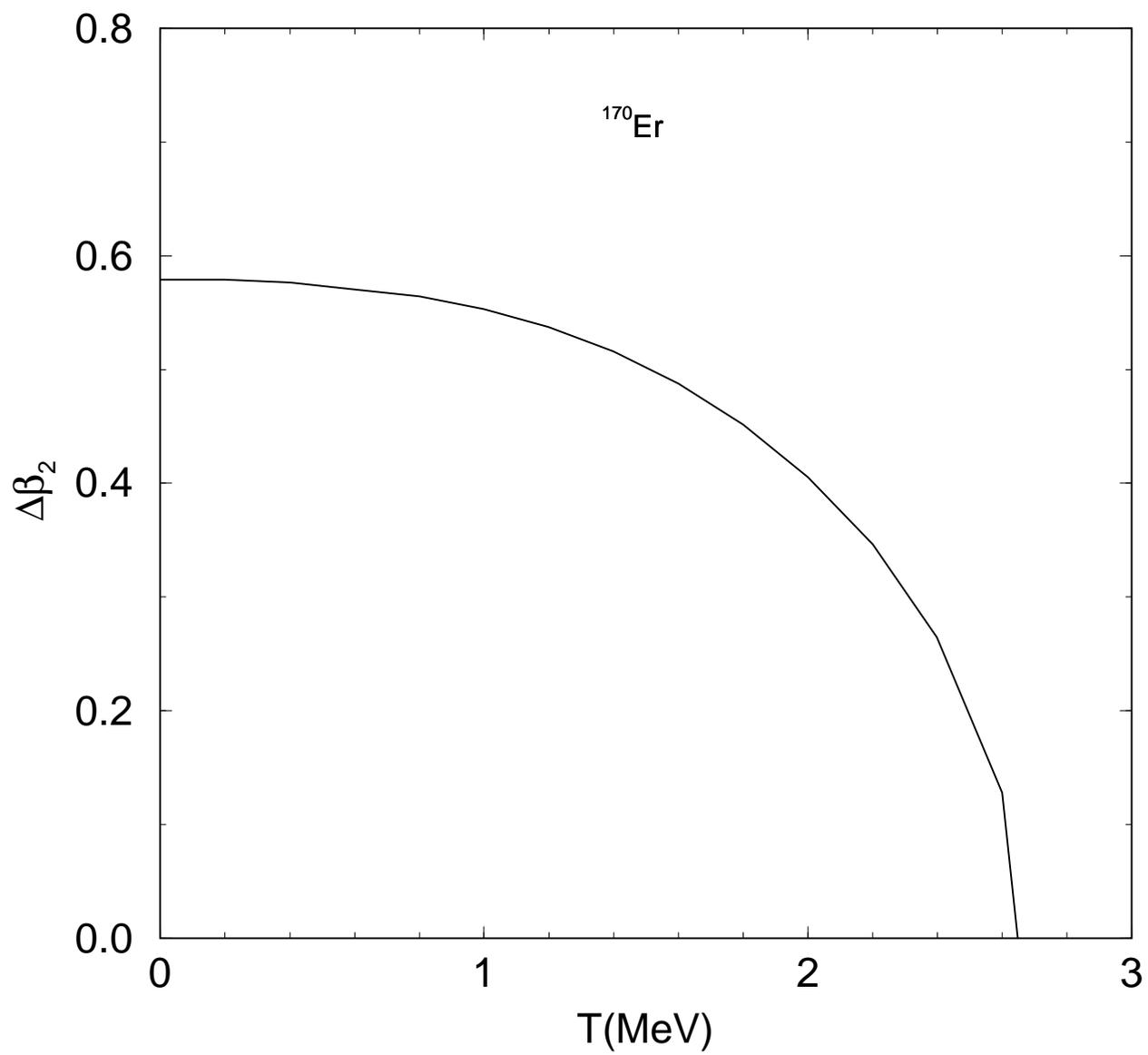
\item[Fig. 9:] The  difference $\Delta \beta_2$ between the deformations 
at the prolate and the oblate
minima as a function of temperature for $^{170}Er$.
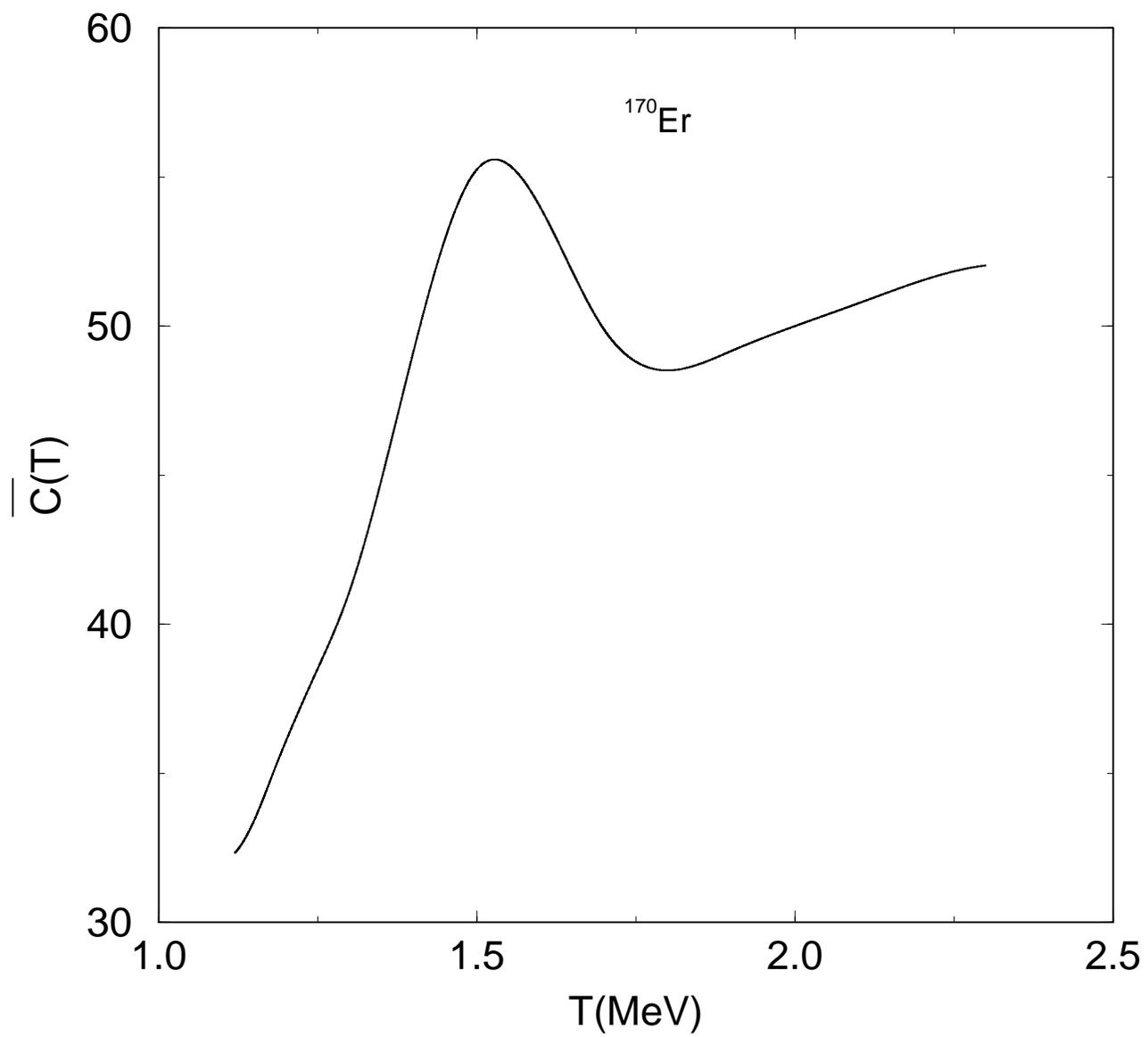
\item[Fig. 10:] Variation of average specific heat  as a function 
of temperature for $^{170}Er$.
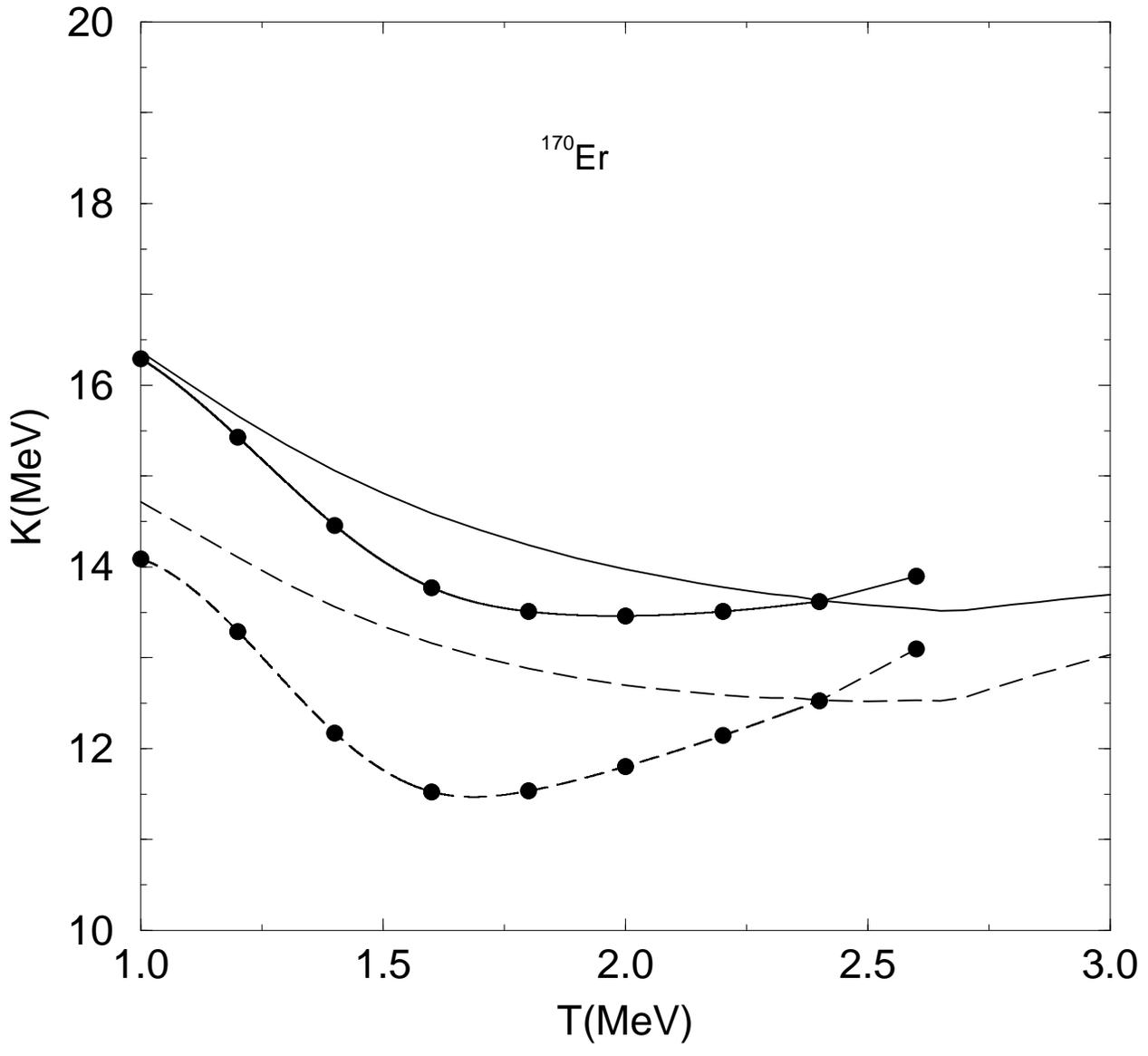
\item[Fig. 11:] The inverse level density parameter for the      
system $^{170}Er$. The full and the dashed lines correspond to
$K_S$ and $K_E$, respectively. The lines with filled circles represent 
the corresponding results with inclusion of thermal fluctuations.
\end{itemize}
\end{document}